# APPLICATION OF THOMAS-FERMI MODEL TO FULLERENE MOLECULE AND NANOTUBE

*UDC 547*

## Yuri Kornyushin

Maître Jean Brunschvig Research Unit Chalet Shalva, Randogne, CH-3975

**Abstract**. *Semiclassical description, based on electrostatics and Thomas-Fermi model is applied here to calculate dimensions of the electronic shell of a fullerene molecule and a nanotube. The internal radius of the electronic shell of a fullerene molecule, calculated within the framework of the model is 0.2808 nm. The external radius is 0.4182 nm. The experimental values are 0.279 nm and 0.429 nm correspondingly. This shows that semiclassical approach provides rather good description of the dimensions of the electronic shell in a fullerene molecule. Two types of dipole oscillations in a fullerene molecule are considered and their frequencies are calculated. Similar calculations are performed for a nanotube also. For a nanotube with a radius of the cylinder of the ions, $R_n = 0.7$ nm, the internal radius of the electronic shell, calculated within the framework of the model is 0.577 nm. The external radius is 0.816 nm. Three types of dipole oscillations in nanotube are considered and their frequencies are calculated.*

### INTRODUCTION

Semiclassical description, based on electrostatics and Thomas-Fermi model was applied rather successfully to study atomic, molecular and nano-objects problems [1-4]. Fullerene molecule was studied in [1,3,4]. Fullerene molecule ($C_{60}$) forms a spherical ball. Let us denote the radius of the sphere, on which the carbon atoms are situated, $R_f$. The value of this radius is assumed to be determined by the chemical bonds. The electronic configuration of the constituent carbon atom is $1s^2 2s^2 2p^2$. It was assumed [1,3,4] that in a fullerene molecule two 1s electrons of each atom belong to the core (forming the ion itself), two 2s electrons form the molecular bonds and two 2p electrons are delocalized, or free. Hence it was assumed [1,3,4] that the total number of the delocalized electrons in a fullerene molecule is 120. Another assumption was made, that the delocalized electrons cannot penetrate inside the sphere of the ions as they repel each other. The latest experimental data [5] show that $R_f = 0.354$ nm, the total number of the delocalized electrons, $N = 240$, the internal radius of the electron shell, $R_i = 0.279$ nm (that is the delocalized electrons do penetrate the sphere of the ions), and the external radius of the electronic shell, $R = 0.429$ nm.

In the model considered it is assumed [1] that the positive charge of the ions, $-eN$ ($e < 0$ is electron charge), is distributed homogeneously on the surface of the sphere $r = R_f$ ($r$ is the distance from the center of the fullerene molecule). The charge of the delocalized electrons, $eN$, is assumed to be distributed homogeneously in a spherical layer $R_i < r < R$ (see Fig. 1).

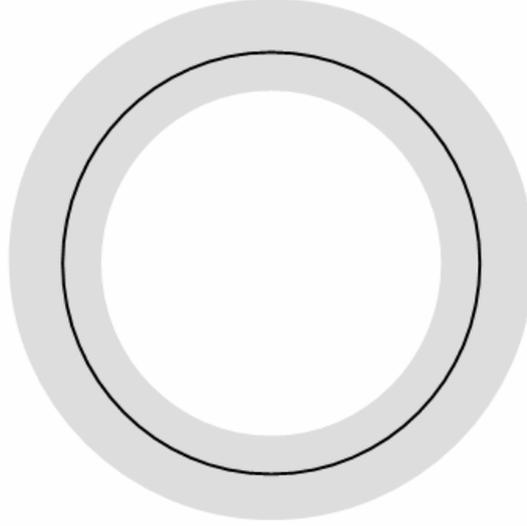

Fig. 1. Model, applied to a fullerene molecule and a nanotube. The charge of the ions is distributed homogeneously on the surface of the sphere of radius $R_f$, or cylinder of radius $R_n$ for a nanotube. The delocalized electrons are distributed uniformly in a spherical (or cylindrical for a nanotube) layer $R_i < r < R$.

Now let us calculate the electrostatic energy of a fullerene molecule.

CALCULATION OF THE ELECTROSTATIC ENERGY OF A FULLERENE MOLECULE

The electrostatic potential $\varphi(\mathbf{r})$, arising from electric charge, can be obtained as a solution of the Poisson's equation [6]

$$\Delta\varphi(\mathbf{r}) = -4\pi\rho(\mathbf{r}), \tag{1}$$

where $\rho(\mathbf{r})$ is the charge density.

The uniformly distributed charge density in the spherical layer of the electronic shell of a fullerene molecule is

$$\rho(\mathbf{r}) = 3eN/4\pi(R^3 - R_i^3). \tag{2}$$

The charge of the delocalized electrons produces the following electrostatic potential inside the spherical layer, where the delocalized electrons are found:

$$\varphi_e = [eN/(R^3 - R_i^3)][(3R^2/2) - (r^2/2) - (R_i^3/r)]. \tag{3}$$

The charge of the ions creates the following potential on the sphere $r = R_f$ and outside of it:

$$\varphi_i(r) = -eN/r. \tag{4}$$

At $r \leq R_f$ the positive charge of the sphere creates electrostatic potential equal to $-eN/R_f$. According to Eqs. (3,4), the total electrostatic potential, $\varphi_e(r) + \varphi_i(r)$, is zero at $r = R$. Hence for $r > R$ it is zero also. Inside the neutral fullerene molecule, when $0 \leq r \leq R_i$, the total electrostatic potential is as follows:

$$\varphi_{in} = [3eN(R + R_i)/2(R^2 + RR_i + R_i^2)] - eN/R_f. \tag{5}$$

As on the external surface of a neutral fullerene molecule, at $r = R$, the total electrostatic potential is zero, it should be zero on the internal surface, at $r = R_i$, also. Otherwise the electrons will move to the locations, where their potential energy is smaller. In a state of equilibrium $\varphi_{in} = 0$. This yields:

$$R_i = 0.5\{R_f - \delta + [R_f^2 - (\delta^2/3)]^{1/2}\}, \tag{6}$$

where $\delta = R - R_i$ is the thickness of the electronic shell. The latest experimental data [5] are: $R_f = 0.354$ nm and $\delta = 0.15$ nm. For these values Eq. (6) yields $R_i = 0.274$ nm. The experimental value is 0.279 nm.

The total electrostatic energy of a system consists of the electrostatic energy of the ions, electrostatic energy of the delocalized electrons, and electrostatic energy of the interaction between them. It is described by the following expression:

$$W = (e^2N^2/2R_f) + (e^2N^2/2R) - [e^2N^2/2R_f(R^3 - R_i^3)](3R_fR^2 - 2R_i^3 - R_f^3)$$

$$+ 0.1[eN/(R^3 - R_i^3)]^2[R^5 + 9R_i^5 - 5(R_i^6/R) - 5R_i^3R^2]. \tag{7}$$

Here the first term describes the electrostatic energy of the ions. The third term describes the interaction energy. The second and the fourth term describe the electrostatic energy of the delocalized electrons. The second term represents the energy of the electrostatic field outside the charged spherical layer. The fourth term represents the energy of the field inside the spherical layer. It is calculated as an integral over the layer of the square of the gradient of $\varphi_e(r)$, divided by $8\pi$.

CALCULATION OF THE KINETIC AND TOTAL ENERGY OF A FULLERENE MOLECULE

The kinetic energy, $T$, of a gas of delocalized electrons, confined in a volume of regarded spherical layer $4\pi(R^3 - R_i^3)/3$, and calculated in accord with the ideas of the Thomas-Fermi model, is [1]

$$T = 1.105(\hbar^2/m)N^{5/3}/(R^3 - R_i^3)^{2/3}. \tag{8}$$

The total energy, $T + W$, is a function of a single variational parameter, $\delta$. Indeed, when Eq. (6) is substituted in Eqs. (7,8) together with $R = R_i + \delta = 0.5\{R_f + \delta + [R_f^2 - (\delta^2/3)]^{1/2}\}$, we see that the total energy is a function of a single variational parameter, $\delta$. In a state of equilibrium the total energy reaches minimum $(T + W)_{min} = 9.422$ keV. The minimum (equilibrium) value of the total energy is reached when $\delta = \delta_e = 0.1374$ nm. At that $R_i = R_{ie} = 0.2808$ nm, and $R = R_e = 0.4182$ nm. The minimum (equilibrium) energy of a positive ion of a fullerene molecule (with $N = 239$) was calculated to be 10.327 keV.

At that in this case $R_i = 0.2808$ nm and $R = 0.4182$. The difference appears in further digits, it is very small.

It is worthwhile noting that the total energy of a fullerene molecule also contains large negative terms, corresponding to the energy of the ionic cores and the energy of the chemical bonds. These terms were not calculated here.

## COLLECTIVE DIPOLE OSCILLATION OF DELOCALIZED ELECTRONS IN A FULLERENE MOLECULE

Two peaks of dipole collective oscillations of delocalized electrons in $C_{60}$ positive ions, discovered experimentally by S. W. J. Scully et al, are reported in [7]. The authors associated the lower peak near 20 eV with a surface plasmon, its frequency being √3 times smaller than the Langmuir frequency. The other peak about 40 eV is associated by the authors with some volume plasmon of unknown origin.

*Mie Oscillation*. Mie oscillation is a collective delocalized electron oscillation in a thin surface layer of a conducting sphere [8,9], where positive and negative charges are not separated in space. So it does not look suitable to regard such an oscillation in the thin surface layer of the electronic shell of a fullerene molecule. Anyway let us estimate possible frequency of such oscillation. Let us consider a thin surface layer of a thickness $h$, situated on the external surface of the electronic shell of a fullerene molecule. The volume of this layer, $4\pi R^2 h$, contains electric charge $4\pi R^2 hen$ (here $n$ is the number of delocalized electrons per unit volume of the electronic shell). Let us shift the layer as a whole by the distance $s$ along the $z$-axis. This shift creates a dipole moment $P = 4\pi R^2 hens$ and (restoring) electric field (in a spherically symmetric object) $E_r = 4\pi P/3 = (4\pi R)^2 hens/3$ correspondingly. Restoring force at that is $(4\pi enR)^2 hs/3$. The mass of the regarded thin surface layer is $4\pi R^2 hnm$ (here $m$ denotes the mass of the delocalized electron). Restoring force leads to the dipole oscillation. From Newton equation follows that the frequency of the oscillation considered is $\omega_M = (4\pi e^2 n/3m)^{1/2}$. This frequency, as was mentioned by S. W. J. Scully et al [7], is √3 times smaller than Langmuir frequency. In the model described above $n = 3N/4\pi(R^3 - R_i^3)$. From two last equations written above follows that $\omega_M = [e^2 N/m(R^3 - R_i^3)]^{1/2}$. This equation yields $\hbar\omega_M = 21.45$ eV when the experimental values of the parameters are used. S. W. J. Scully et al reported the value of one of the two measured peaks, $\hbar\omega_M$, near 20 eV [7]. The agreement is quite reasonable. Using theoretical values of $R = 0.4182$ nm and $R_i = 0.2808$ nm, calculated here, one can obtain the value of the peak 22.72 eV. This result is not too bad either.

*Simple Dipole Oscillation*. Let us consider a dipole type collective quantum oscillation of the electronic shell as a whole (having mass $mN$) relative to the ion skeleton. Such an oscillation looks more plausible in a fullerene molecule than Mie oscillation. The value of the electrostatic potential inside the electronic shell is $\varphi_e = [eN/(R^3 - R_i^3)][(3R^2/2) - (r^2/2) - (R_i^3/r)]$ (see Eq. (3)). Let us consider the shift of the electronic shell by a small distance, **s**, relative to the ion skeleton and calculate the change in the electrostatic energy of a fullerene molecule (in the framework of the accepted model). The first term in Eq. (3) does not contribute to the change in the electrostatic energy, as it is constant. The contribution of the third term is also zero because the third term depends on the distance like $(1/r)$, a potential produced by a point charge. Interaction of the ion skeleton with the point charge, situated inside it does not depend on their relative position, because the potential inside the uniformly charged sphere is constant. The change in the electrostatic energy does not depend on the direction of the shift (**s** vector) because of the spherical symmetry of the problem. This means that the change cannot contain term, proportional to $s$. Let us choose the $x$-axis along the **s** vector. Then after the shift we shall have in the third term of Eq. (3) $[(x - s)^2 + y^2 + z^2]$ instead of $r^2 = (x^2 + y^2 + z^2)$. Taking into account that the term, proportional to $s$ in the change of the electrostatic energy is zero, we find that the third term (the only one, which contributes) yields the following change in the electrostatic energy of a fullerene molecule due to the shift regarded: $W(s) = (eNs)^2/2(R^3 - R_i^3)$. This energy, $W(s)$, is a potential of a 3-d harmonic oscillator [10]. For a 3-d harmonic oscillator (mass $mN$) $W(s) \equiv mN\omega_s^2 s^2/2$ [10]. As follows from the last two equations, the frequency, $\omega_s$, is equal to Mie frequency, $\omega_M$. The distance between the quantum levels is $\hbar\omega_M$. Its calculated value is 21.45 eV. Its experimental value is about 20 eV. Quantum excitation to the first excited level manifests itself as 20 eV peak, to the second one as 40 eV peak. Conception of some volume plasmon of unknown origin [7] is not needed.

Mie oscillation is also a 3-d (quantum) harmonic oscillator. So the results are the same for both types of oscillations. Only Mie oscillation in a fullerene molecule looks less plausible.

The frequency of the simple dipole oscillation is equal to the Mie frequency in spherically symmetric objects [1]. So it is rather difficult to determine which type of the oscillation occurs in the object under investigation.

### CALCULATION OF THE ELECTROSTATIC ENERGY OF A NANOTUBE

Let us consider a long (comparative to its diameter) nanotube. Let us denote $N_n$ the number of delocalized electrons in a nanotube per unit length; $R_n$ denotes the radius of the cylinder, on which the ions of a nanotube are situated.

The Gauss theorem [6] allows calculating the electrostatic field inside the electronic shell of a long nanotube (at $R_i \leq r \leq R$):

$$E_e(r) = (2eN_n/r)(r^2 - R_i^2)/(R^2 - R_i^2). \tag{9}$$

The electrostatic field, produced by the ions at $R_n \leq r \leq R$ can be calculated also:

$$E_i(r) = -2eN_n/r. \tag{10}$$

At $r < R_n$ the electrostatic field, produced by the ions, is zero. The electrostatic potential difference between the internal and external surfaces of the electronic shell in a nanotube should be zero in a state of equilibrium. Otherwise the delocalized electrons will move to the locations, where their potential energy is smaller. The electrostatic potential difference is an integral of the electrostatic field between the surfaces mentioned. It is zero when

$$R_i = R^2/R_n \exp 0.5. \tag{11}$$

Eq. (11) yields for the thickness of the electronic shell, $\delta = R - R_i$, the following result:

$$\delta = (R_i R_n)^{1/2}(\exp 0.25) - R_i. \tag{12}$$

As $\delta = R - R_i$, it follows from Eq. (12) that

$$R = R_i + \delta = (R_i R_n)^{1/2} \exp 0.25. \tag{13}$$

The total electrostatic energy of a system is equal to the integral of the square of the electrostatic field, divided by $8\pi$. The value of it per unit length is described by the following expression:

$$W = [e^2 N_n^2/(R^2 - R_i^2)^2][R^4 \ln(R/R_n) + R_i^4 \ln(R_n/R_i)]$$
$$+ [e^2 N_n^2/(R^2 - R_i^2)](R_n^2 - 0.75R^2 - 0.75R_i^2). \quad (14)$$

Here (see Eq. (13)), $R = (R_i R_n)^{1/2} \exp 0.25$. From this follows that the electrostatic energy of the system is a function of only one variational parameter, $R_i$.

### CALCULATION OF THE KINETIC AND TOTAL ENERGIES OF A NANOTUBE

The kinetic energy per unit length, $T$, of a gas of delocalized electrons, confined in a cylindrical layer of a volume per unit length $\pi(R^2 - R_i^2)$, and calculated in accord with the ideas of the Thomas-Fermi model, is as follows:

$$T = 1.338(\hbar^2/m)N_n^{5/3}/R_i^{2/3}[R_n(\exp 0.5) - R_i]^{2/3}. \quad (15)$$

The kinetic energy is a function of only one variational parameter, $R_i$, Thence, the total energy, $T + W$, is a function of a single variational parameter, $R_i$. In a state of equilibrium the total energy reaches a minimum $(T + W)_{min} = 71.51$ keV/nm at $R_n = 0.7$ nm and $N_n = 670$ 1/nm. Parameter $R_n = 0.7$ nm was taken from [4]. It was assumed here also that the total number of delocalized electrons in a nanotube is four times larger than the number of carbon atoms. For the same values of $R_n = 0.7$ nm and $N_n = 670$ 1/nm the minimum (equilibrium) value of total energy is reached when $R_i = R_{ie} = 0.577$ nm. At that $R = R_e = 0.816$ nm and $\delta_e = 0.239$ nm.

It is worthwhile noting that the total energy of a nanotube also contains large negative terms, corresponding to the energy of the ionic cores and the energy of the chemical bonds. These terms were not calculated here.

### COLLECTIVE DIPOLE OSCILLATIONS OF DELOCALIZED ELECTRONS IN A NANOTUBE

Longitudinal surface dipole collective oscillation of delocalized electrons in a nanotube was first described in [2]. The frequency of this oscillation is [2]

$$\omega_l = e(\pi N_n/lmR)^{1/2}, \quad (16)$$

where $l$ is the length of a nanotube.

At $N_n = 670$ 1/nm, $l = 30$ nm, $R = 0.816$ nm we have $\hbar\omega_l = 3.065$ eV.

Surface dipole collective oscillation of delocalized electrons in a conductive sphere is called, as was mentioned above, Mie oscillation [8,9]. The frequency of the Mie oscillation is square root from 3 times smaller than the Langmuir frequency [2], $\omega_L = (4\pi e^2 n/m)^{1/2}$ (here $n$ is the number of the delocalized electrons per unit volume of the electronic shell). In general case [9] the frequency of Mie oscillation, $\omega_M = (4\pi D e^2 n/m)^{1/2}$ (here $D$ is the depolarization factor [9]). Since for a sphere [6] $D = 1/3$, we have for Mie oscillation in a sphere $\omega_M = (4\pi e^2 n/3m)^{1/2}$, as was mentioned above. For a cylinder [6] $D = 1/2$ and $n = N_n/\pi(R^2 - R_i^2)$. Thence we have for the frequency of the transverse surface (Mie) oscillation in nanotube following result:

$$\omega_M = [2e^2 N_n / m(R^2 - R_i^2)]. \tag{17}$$

For $N_n$ = 670 1/nm, $R$ = 0.816 nm, and $R_i$ = 0.577 nm we have $\hbar\omega_M$ = 21 eV.

Like for the fullerene molecule, when the electronic shell of a nanotube is shifted as a whole by a small distance, **s**, relative to the ion skeleton, the electrostatic energy of the nanotube is changed by $W(s)$, which is a potential of a 2-d harmonic oscillator. Let us calculate this change. The electrostatic potential is equal to the integral of minus electric field [6]. Inside the electronic shell of a nanotube, as follows from Eq. (9) the electrostatic potential is as follows:

$$\varphi_e = -[eN_n r^2/(R^2 - R_i^2)] - 2eN_n[R_i^2/(R^2 - R_i^2)]\ln(r/r_0), \tag{18}$$

where $r_0$ is some constant. The second term in Eq. (18) coincides with the potential produced by a linear charge. This term does not contribute to the change in the electrostatic energy of a nanotube, because the interaction between the ion skeleton and a linear charge is zero. This is so because the electrostatic potential inside the uniformly charged cylinder is constant. The change in the electrostatic energy does not contain a term, proportional to $s$ because of the radial symmetry of the problem. So the only contribution comes from the first term of Eq. (18). Let us choose the $x$-axis along the **s** vector. Then after the shift we shall have in the first term of Eq. (18) $[(x - s)^2 + y^2]$ instead of $r^2 = (x^2 + y^2)$. Taking into account that the term, proportional to $s$ in the change of the electrostatic energy is zero, we find that the first term (the only one, which contributes) yields the following value of the change in the electrostatic energy of a nanotube (per unit length) due to the shift regarded: $W(s) = (eN_n s)^2/(R^2 - R_i^2)$ [10]. This energy, $W(s)$, is a potential of a 2-d harmonic oscillator [10]. For a 2-d harmonic oscillator [10] (mass per unit length of a nanotube $mN_n$) $W(s) \equiv mN_n \omega_s^2 s^2/2$. As follows from the last two equations, the frequency, $\omega_s$, is equal to Mie frequency, $\omega_M$. The distance between the levels is $\hbar\omega_M$. Its calculated value is 21 eV. So, according to the calculations performed, the quantum transition to the first excited level should manifest itself as 21 eV peak, and to the second one as 42 eV peak.

Mie oscillation in a nanotube is also a 2-d harmonic oscillator. So the results are the same for both types of the oscillations. Only Mie oscillation in a fullerene molecule looks less plausible.

Obtained results show that a very simple semiclassical concept of Thomas-Fermi model often gives rather good agreement between experimental and theoretical results. In particular, this model often gives rather reasonable description of nanoobjects.

## DISCUSSION

Today many authors use the model of the electronic shell accepted here. This model allows understanding experimental results. As a matter of fact this concept is present in papers [5,7]. Two peaks of dipole oscillations of the delocalized electrons in $C_{60}$ ions, discovered experimentally by S. W. J. Scully et al, are reported [7]. The authors ascribe these peaks to the surface (Mie) oscillation [8,9] and some bulk dipole mode. The energy values of the quanta of the two peaks were measured to be about 20 eV and 40 eV. Using experimental data on $N$, $R$ and $R_i$ and equation for the Mie frequency [2,9], one could calculate the energy of the quantum of the Mie oscillation, 21.45 eV.

Oscillation of the electronic shell as a whole relative to the skeleton of the ions is an oscillation of a 3-d (for a fullerene molecule) and 2-d (for a nanotube) harmonic oscillator. The same is for the Mie oscillation. The cited experimental data could be explained this way also.

Obtained results show that a very simple semiclassical concept, based on electrostatics and Thomas-Fermi model often gives rather good agreement between experimental and theoretical results. In particular, this model gives rather reasonable description of nanoobjects.

# PRIMENA THOMAS-FERMIJEVOG MODELA NA MOLEKULE FULERENA I NANOTUBE

## Yuri Kornyushin

*U ovom radu su primenom poluklasičnog pristupa zasnovanog na elektrostatičkom i Thomas-Fermijevom modelu izračunate dimenzije elektronske ljuske molekula fulerena i nanotuba. Izračunata vrednost unutrašnjeg poluprečnika elektronske ljuske molekula fulerena je 0.2808 nm, a spoljašnjeg 0.4182 nm. Odgovarajuće eksperimentalno određene vrednosti su 0.279 nm i 0.429 nm. Ovo slaganje pokazuje da poluklasičan pristup dobro opisuje veličinu elektronske ljuske molekula fulerena. Razmatrana su i dva tipa dipolnih oscilacija u tim molekulima i izračunate su njihove frekvencije. Slični proračuni su primenjeni i za nanotubu. Za nanotubu sa poluprečnikom jonskog cilindra Rn = 0.7 nm, unutrašnji poluprečnik elektronske ljuske izračunat na osnovu ovog modela je 0.577 nm, a spoljašni 0.816 nm. Takođe su razmatrana i tri tipa dipolnih oscilacija nanotube i izračunate su njihove frekvencije.*